\documentclass[10pt]{article}
\usepackage[utf8]{inputenc}
\usepackage{graphicx}
\usepackage{dcolumn}
\usepackage{bm}
\usepackage{chngpage}
\usepackage{graphicx}
\usepackage{braket}
\usepackage[margin=0.9in]{geometry}
\usepackage{amsmath}
\usepackage{float}
\usepackage{graphicx}
\usepackage{caption}
\usepackage{amsmath, amssymb}
\usepackage{bbold}
\usepackage{makecell}
\usepackage{braket}
\usepackage{calc}
\usepackage{subcaption}
\usepackage{authblk}
\usepackage{indentfirst}
\usepackage{lipsum}
\usepackage{marvosym}
\usepackage{lineno}
\usepackage{soul}
\usepackage[version=4]{mhchem}

\usepackage[
autocite=superscript,
backend=biber,
style=nature,
date=year,
doi=false,isbn=false,url=false,eprint=false
]{biblatex}
\usepackage[lang=en]{jabbrv}
\AtEveryBibitem{\clearlist{language}}
\usepackage[labelfont=bf]{caption}
\usepackage{amsmath}
\usepackage[dvipsnames]{xcolor}
\usepackage{geometry}
 \usepackage[%
    colorlinks=true,
    pdfborder={0 0 0}{},
    linkcolor=blue,
    citecolor=blue,
]{hyperref}
\usepackage{textcomp, gensymb}
\DefineJournalPartialAbbreviation{Rhapsod}{Rhaps}

\title{\textbf{Terahertz Control of Linear and Nonlinear Magno-Phononics}}
\author[1,*]{Tianchuang Luo}
\author[1,*]{Honglie Ning}
\author[1,*]{Batyr Ilyas}
\author[1,*]{Alexander von Hoegen}
\author[2,3]{Emil Viñas Boström}
\author[4]{Jaena Park}
\author[4]{Junghyun Kim}
\author[4]{Je-Geun Park}
\author[5]{Dominik M. Juraschek}
\author[2,6]{Angel Rubio}
\author[1,\Letter]{Nuh Gedik}

\affil[1]{Department of Physics, Massachusetts Institute of Technology, Cambridge, 02139, Massachusetts, USA.}
\affil[2]{Max Planck Institute for the Structure and Dynamics of Matter, Luruper Chaussee 149, 22761 Hamburg, Germany.}
\affil[3]{Nano-Bio Spectroscopy Group, Departamento de Fisica de Materiales, Universidad del Pais Vasco, 20018 San Sebastian, Spain.}
\affil[4]{Department of Physics and Astronomy and Institute of Applied Physics, Seoul National University, Seoul 08826, Republic of Korea.}
\affil[5]{School of Physics and Astronomy, Tel Aviv University, Tel Aviv 6997801, Israel.}
\affil[6]{Center for Computational Quantum Physics, The Flatiron Institute, New York, NY 10010, USA.}
\affil[\Letter]{e-mail: gedik@mit.edu}
\affil[*]{These authors contributed equally to this work.}

\begin{document}
\maketitle

\section*{Abstract}
Coherent manipulation of magnetism through the lattice provides unprecedented opportunities for controlling spintronic functionalities on the ultrafast timescale. Such nonthermal control conventionally involves nonlinear excitation of Raman-active phonons which are coupled to the magnetic order. Linear excitation, in contrast, holds potential for more efficient and selective modulation of magnetic properties. However, the linear channel remains uncharted, since it is conventionally considered forbidden in inversion symmetric quantum materials. Here, we harness strong coupling between magnons and Raman-active phonons to achieve both linear and quadratic excitation regimes of magnon-polarons, magnon-phonon hybrid quasiparticles. We demonstrate this by driving magnon-polarons with an intense terahertz pulse in the van der Waals antiferromagnet $\mathrm{FePS_3}$. Such excitation behavior enables a unique way to coherently control the amplitude of magnon-polaron oscillations by tuning the terahertz field strength and its polarization. The polarimetry of the resulting coherent oscillation amplitude breaks the crystallographic $C_2$ symmetry due to strong interference between different excitation channels. Our findings unlock a wide range of possibilities to manipulate material properties, including modulation of exchange interactions by phonon-Floquet engineering.
\section*{Main Text}

\begin{center}
\begin{figure}[H]
   \sbox0{\includegraphics[width=\textwidth]{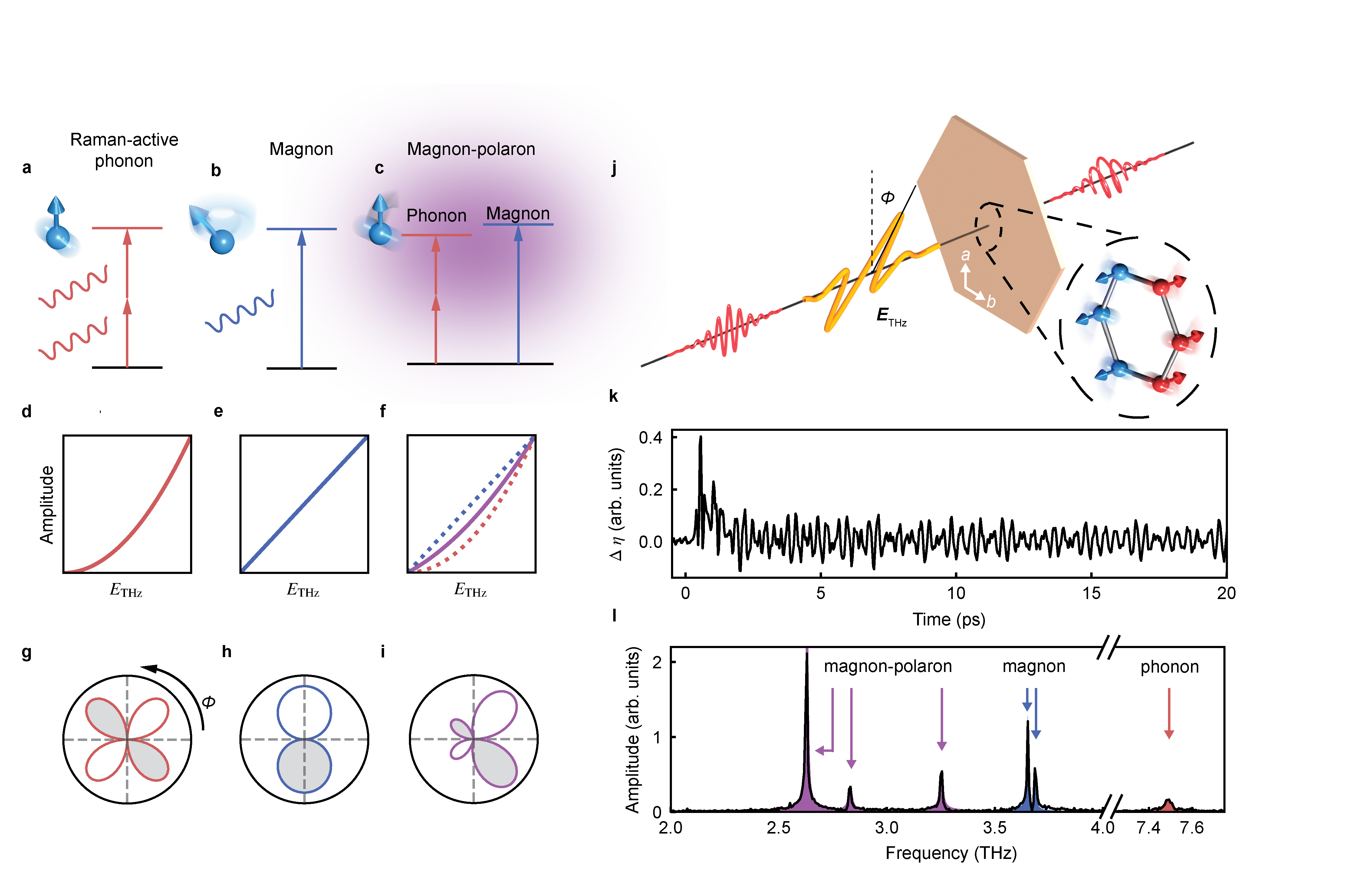}}
    \begin{minipage}{\wd0}
  \usebox0
  \captionsetup{labelfont={bf},labelformat={default},labelsep=period,name={Fig.}}
  \caption{\textbf{Nonlinear and linear excitation pathways.} \textbf{a-c}, Schematics of the excitation mechanisms for a Raman-active phonon, magnon, and magnon-polaron (MP). \textbf{d-f}, Driving field $E_\mathrm{THz}$-dependence of the amplitudes of the modes shown in panels \textbf{a-c}. The Raman-active phonon shows a quadratic dependence (\textbf{d)}. The magnon shows a linear dependence (\textbf{e}). The MP shows a dependence described by Eq.~\ref{fluence} (violet line in \textbf{f}), a combination of quadratic (dotted red line in \textbf{f}) and linear (dotted blue lien in \textbf{f}) dependence. \textbf{g-i}, Driving field polarization $\phi$-dependence of the amplitudes of different modes. Filled and white lobes denote opposite phases. \textbf{g}, The Raman-active phonon exhibits a four-petal pattern described by a second-order homogeneous polynomial of $\sin\phi$ and $\cos\phi$. \textbf{h}, The magnon exhibits a two-petal pattern described by a linear homogeneous polynomial of $\sin\phi$ and $\cos\phi$. \textbf{i}, The MP exhibits a pattern breaking the $C_2$ symmetry as a result of the combination of nonlinear and linear excitation pathways. \textbf{j}, Schematic of the experimental setup. A strong THz pump (yellow) and a near-IR probe (red) are focused on the (001) surface of $\mathrm{FePS_3}$ with a tunable time delay. Both $E_\mathrm{THz}$ and $\phi$ can be controlled (Methods and Supplementary Note 2). Different modes in $\mathrm{FePS_3}$ (bottom right) are detected through the change in linear birefringence ($\Delta$LB) of the transmitted probe. \textbf{k}, $\Delta$LB transient obtained at 10 K at maximal $E_\mathrm{THz}$ and $\phi=52^\circ$. \textbf{l}, FFT spectrum of the time trace in \textbf{k}, where MPs, magnons, and Raman-active phonons are indicated by violet, blue, and red arrows, respectively. The shaded regions are fits to the FFT peaks by damped oscillators.} 
  \label{Fig1}
\end{minipage}
\end{figure}  
\end{center}

\paragraph{}
Coupling between various degrees of freedom in quantum materials gives rise to exotic phases and powerful controls over functional properties. Exploiting the lattice degree of freedom in magnetic materials offers unique perspectives into emerging ultrafast spintronic and straintronic applications\autocite{Kirilyuk2010UltrafastOrder,Nemec2018AntiferromagneticOpto-spintronics}{}. By coherently driving phonons in the terahertz (THz) spectral range, an ultrafast and nonthermal control of magnetism circumventing laser-induced heating can be realized. It has been experimentally demonstrated that coherent phonons can impact the magnetic degree of freedom through a large variety of pathways, including facilitating ultrafast demagnetization\autocite{Maehrlein2018DissectingExcitation,Forst2011DrivingExcitation,Forst2015SpatiallyHeterointerface}{}, driving spin-wave precession\autocite{Nova2017AnPhonons,Rongione2023EmissionInteractions}{}, modulating exchange interactions and magnetic anisotropy \autocite{Melnikov2003CoherentSurface,Kim2012UltrafastFilms,Deb2019FemtosecondGarnets,Soumah2021OpticalCoupling}{}, inducing dynamical magnetization\autocite{Juraschek2020Phono-magneticEffects,Juraschek2022GiantParamagnets,Luo2023LargeHalides,Davies2024PhononicEffect,Basini2024TerahertzSrTiO3}, mediating angular momentum transfer\autocite{Dornes2019TheEffect,Tauchert2022PolarizedDemagnetization}{}, flipping magnetic order parameter\autocite{Stupakiewicz2021UltrafastMagnetization,Gidding2023DynamicMagnon-polarons}{}, stabilizing fluctuating magnetization\autocite{Disa2023Photo-inducedYTiO3}{}, and inducing exotic magnetic states inaccessible via equilibrium approaches\autocite{Disa2020PolarizingField,Afanasiev2021UltrafastPhonons,Kim2012UltrafastVibrations}{}. Conventionally, such control has been achieved through the displacive or impulsive excitation of Raman-active phonons or rectifications to the microscopic parameters with Raman symmetry, whose amplitudes both scale quadratically with the electric field component of light (Fig.~\ref{Fig1}a, see also Supplementary Note 1). Modulating the magnetic properties via linear excitation of Raman-active phonons has not been widely studied. This is because such excitation channel is conventionally considered forbidden due to the absence of an electric-dipole moment of Raman-active phonons in systems with inversion symmetry. Confining excitation exclusively to nonlinear pathways may constrain the attainable amplitude of Raman-active phonons and hinders a detailed control over phonon oscillation patterns, thereby limiting the scope of phononic control of magnetic properties. 

\paragraph{}
A promising avenue to overcome this obstacle involves leveraging Raman-active phonon modes that are linearly coupled to magnons. Since magnons can be linearly excited by magnetic field through the magnetic dipolar process\autocite{Kampfrath2011CoherentWaves,Mashkovich2021TerahertzLattice}{} (Fig.~\ref{Fig1}b), its linear coupling to a Raman-active phonon imparts the phonon with a magnetic dipole, thereby offering a linear pathway for exciting the Raman-active phonon using a resonant magnetic field. The strength of the magnetic-dipolar pathway can be comparable to or even stronger strength than that of the quadratic Raman process. The hybrid magnon-phonon mode, denoted as magnon-polaron (MP), can arise when the frequencies of the two quanta are in vicinity to each other in a system with strong spin-lattice coupling. MPs exhibit the characteristics of both a Raman-active phonon and a magnon (Fig.~\ref{Fig1}c), holding the potential of both nonlinear and linear manipulations. However, despite its promise, a coherent control of the two pathways, including tuning their interference and switching their relative weights, has remained an elusive goal.

\paragraph{}
Previous investigations on MPs largely focused on the acoustic phonon-magnon hybrid in ferromagnetic materials, where the coupling is prominent at finite momentum and in gigahertz range\autocite{Temnov2012UltrafastAcousto-magneto-plasmonics,Jager2015ResonantPhonons,Kikkawa2016MagnonEffect,Man2017DirectGarnet,Holanda2018DetectingExperiments,Scherbakov2010CoherentPulses,Zhang2020High-frequencyMultilayers,Godejohann2020MagnonFerromagnet,Liu2021SignatureGadolinium}{}. Here, we demonstrate both linear and nonlinear coherent excitation of MPs in a quasi-two-dimensional van der Waals antiferromagnet $\mathrm{FePS_3}$, which exhibits strong spin-lattice coupling\autocite{Ergecen2023CoherentAntiferromagnet,Zhang2021SpinFePS3,Zhou2022DynamicalAntiferromagnets,Zong2023Spin-mediatedAntiferromagnet,Mertens2023UltrafastFePS3}
{}. \ce{FePS3} respects 2/m point group symmetry and orders into a zigzag antiferromagnetic pattern with an easy axis along the out-of-plane direction below 118 K (Fig.~\ref{Fig1}j inset)\autocite{Kurosawa1983NeutronFePS3}. Importantly, the magnon frequency of $\mathrm{FePS_3}$ lies in proximity to the Raman-active phonon frequencies, enabling strong hybridization as evidenced by the avoided crossing between them in the THz range\autocite{Zhang2021CoherentInsulator,Vaclavkova2021MagnonFePS3,McCreary2020Quasi-two-dimensionalSpectroscopy,Liu2021DirectFields}{}. Here, we use high-field, ultrashort THz electromagnetic pulses that possess significant spectral weight at the MP frequencies to resonantly excite the MPs and enable their linear excitation. Meanwhile, the THz pulses are tailored to cover a broad bandwidth to facilitate the nonlinear Raman excitation pathway (Supplementary Note 2).

\paragraph{}
To distinguish between the linear and nonlinear pathways, we analyze their distinct dependencies on the THz electric field strength $E_\mathrm{THz}$ and THz electric field polarization angle $\phi$ (Fig.~\ref{Fig1}j). The amplitude ($A$) of a mode driven by the Raman excitation channel follows a quadratic dependence on $E_\mathrm{THz}$ (Fig.~\ref{Fig1}d), whereas it exhibits a linear dependence on $E_\mathrm{THz}$ for the resonant pathway (Fig.~\ref{Fig1}e).  The mode amplitude is then determined by the sum of their contributions (Fig.~\ref{Fig1}f): 
\begin{equation}
\label{fluence}
A=|aE_\mathrm{THz}^2+bE_\mathrm{THz}|,
\end{equation}
where $a$ and $b$ represent the excitation efficiencies of the nonlinear and linear pathways, respectively.
\paragraph{}
Additionally, the amplitude exhibits a distinct behavior as a function of the THz polarization angle $\phi$ (see Fig. \ref{Fig1}g-i). In the linear pathway, the magnetic component of the THz driving field directly interacts with the magnetic moment induced by magnons and can thus linearly excite them. The MP amplitude is maximal when the magnetic field is aligned with the magnetization generated by the magnon. Therefore, the $\phi$-dependence of the MP amplitude follows a two-petal pattern represented by a linear polynomial of $\cos(\phi)$ or $\sin(\phi)$ (Fig.~\ref{Fig1}h). On the other hand, the nonlinear Raman excitation pathway follows a homogeneous quadratic polynomial of $\cos(\phi)$ and $\sin(\phi)$ (Fig.~\ref{Fig1}g), with the specific form determined by the symmetry of the mode\autocite{Mashkovich2021TerahertzLattice,Johnson2019DistinguishingSpectroscopy}{}. The phase of the mode driven by the two channels exhibit different $\phi$ dependencies (Fig.~\ref{Fig1}g-h shading of petals). The interference between these two routes will lead to a polar pattern with unequal amplitude when the direction of the driving field is flipped ($\phi\rightarrow\phi+180^\circ$), breaking the $C_2$ symmetry along the $b$-axis of $\mathrm{FePS_3}$ (Fig.~\ref{Fig1}i). Therefore, by scanning $E_\mathrm{THz}$ and $\phi$, we can differentiate the two excitation methods. More importantly, we gain control of their relative strengths and can therefore coherently shape the symmetry of the MP dynamics.

\paragraph{}
In our experiment, we measure the MPs by tracking the polarization state of an ultrashort, 800 nm probe pulse that passes through the sample. By controlling the relative delay $\Delta t$ between the probe pulse and the THz excitation, the MPs appear as coherent oscillations in the probe pulse ellipticity ($\Delta\eta$). Fig.~\ref{Fig1}k shows a typical $\Delta\eta$ transient measured in $\mathrm{FePS_3}$ at 10 K as a function of $\Delta t$. The initial positive signal close to $\Delta t = 0$ arises from the THz Kerr effect when pump and probe temporally overlap. A beating pattern emerges at later delays, which indicates the coexistence of multiple coherent oscillations. A fast Fourier transform (FFT) of the time trace in Fig.~\ref{Fig1}k reveals a series of peaks corresponding to various low-energy excitations of the system (Fig.~\ref{Fig1}l). The three lowest energy modes (violet shadings) have been shown to hybridize with the magnons (blue shadings) under magnetic field\autocite{Zhang2021CoherentInsulator,Vaclavkova2021MagnonFePS3,McCreary2020Quasi-two-dimensionalSpectroscopy,Liu2021DirectFields}{}, indicating that they should be referred to as MPs instead of pure phonons, as will be shown unambiguously in the following.

\begin{center}
\begin{figure}[H]
   \sbox0{\includegraphics{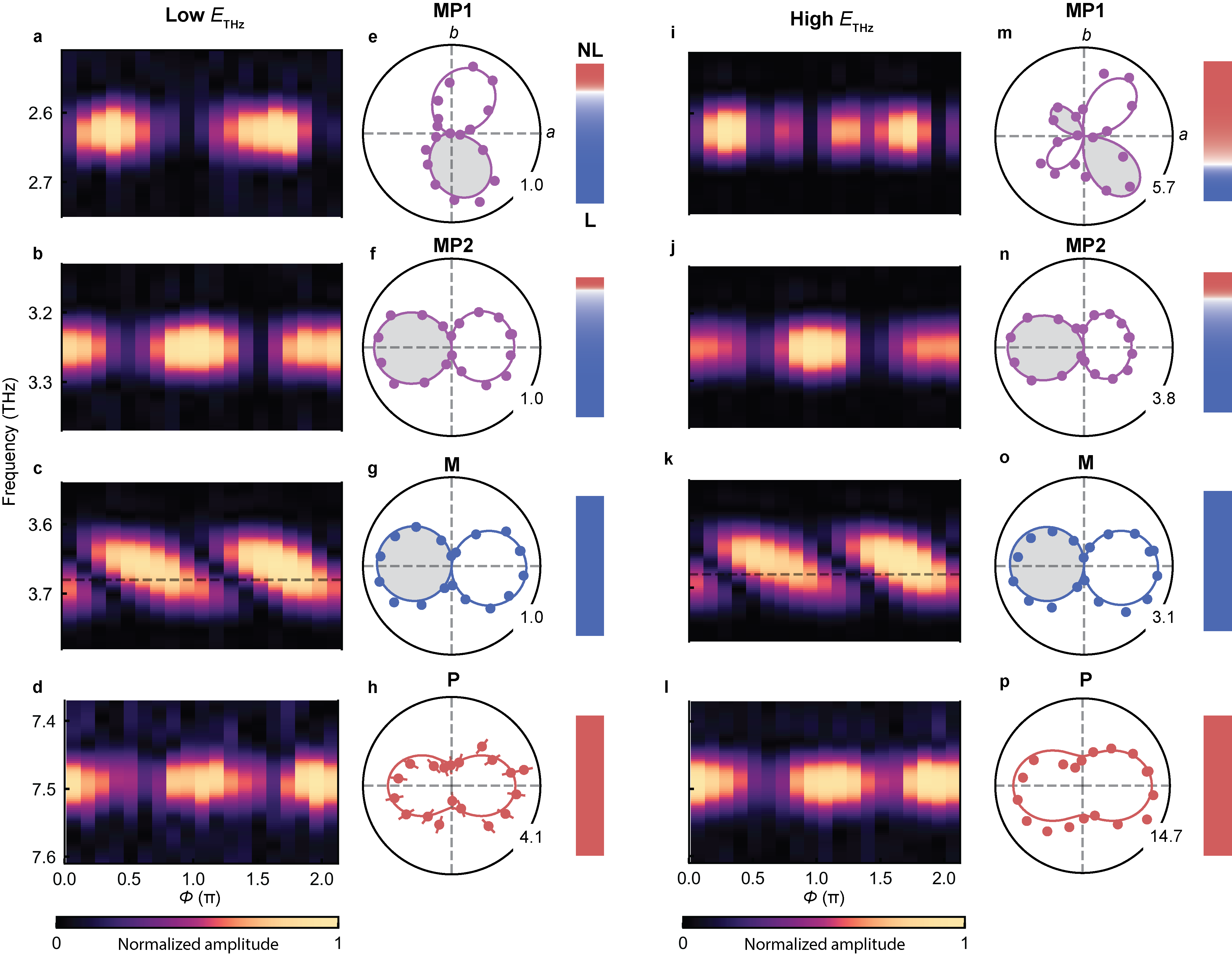}}
    \begin{minipage}{\wd0}
  \usebox0
  \captionsetup{labelfont={bf},labelformat={default},labelsep=period,name={Fig.}}
  \caption{\textbf{$\phi$ dependence of the mode amplitudes.} \textbf{a-d}, FFT spectra of MP1, MP2, M, and P as a function of THz polarization $\phi$ at low $E_\mathrm{THz}$. The amplitudes are normalized to the maximal value for each mode (see colorbar at the bottom). The horizontal dashed line at 3.68 THz marks the frequency of M. The MP1, MP2, and M modes are measured at 30\% of the maximal $E_\mathrm{THz}$ and the P mode is measured at 50\% of the maximal $E_\mathrm{THz}$. $E_\mathrm{THz}=150 \text{kV/cm}$ \textbf{e-h}, Amplitudes of MP1 (\textbf{e}), MP2 (\textbf{f}), M (\textbf{g}), and P (\textbf{h}) as a function of $\phi$ extracted from \textbf{a-d}. Solid lines are fits. MP1 is fit by the sum of Eq.~\ref{nonlinear} and Eq.~\ref{linear} for a $B_g$ mode. MP2 is fit by the sum of Eq.~\ref{nonlinear} and Eq.~\ref{linear} for a $A_g$ mode. M is fit by Eq.~\ref{linear} for a $A_g$ mode. P is fit by Eq.~\ref{nonlinear} for a $A_g$ mode. The numbers on the polar plots represent the maximal amplitudes normalized to the maximal amplitude measured at 30\% of the maximal $E_\mathrm{THz}$ of different modes. The color bars next to the polar plots indicate the relative strengths of the nonlinear (red) and linear (blue) pathways obtained by the fitting. \textbf{i-p} are identical to panels \textbf{a-h} but acquired at 100\% of the maximal $E_\mathrm{THz}$. The chosen $\phi$ values are not perfect aligned with the crystallographic axes due to difficulties in sample alignment.} 
  \label{Fig3}
\end{minipage}
\end{figure}  
\end{center}

\paragraph{}
To gain insight into the symmetry and the excitation mechanisms of these modes, we investigate their amplitudes as a function of THz polarization $\phi$\autocite{Zhang2024TerahertzAntiferromagnet,Zhang2024Terahertz-field-drivenAntiferromagnet}. We rotate the THz polarization $\phi$ while keeping $E_\mathrm{THz}$, the probe polarization, and the sample orientation unchanged. Mode amplitudes are determined by integrating the area under their respective FFT peaks or by fitting to damped harmonic oscillators (Supplementary Note 3). The excitation spectrum of the modes at different $\phi$ are shown by the color plot in Fig.~\ref{Fig3}. Here we focus on four modes: the 2.7 THz mode (MP1), 3.3 THz mode (MP2), the mode with higher frequency around 3.7 THz (M), and the 7.5 THz mode (P). Further details regarding the remaining modes are provided in Supplementary Notes 4 and 5. We first measure the polarimetry of different modes at a low $E_\mathrm{THz}$ ($\sim$50 kV/cm peak electric field). Both MP1 and MP2 show distorted two-petal patterns breaking the $C_2$ symmetry along the $b$-axis (Figs.~\ref{Fig3}a-b). For MP1, the two lobes are rotated towards $\phi=\pm 45^\circ$ (Fig.~\ref{Fig3}e), while for MP2, the left lobe is noticeably larger than the right lobe (Fig.~\ref{Fig3}f). In comparison, M shows two identical lobes (Figs.~\ref{Fig3}c,g), and P exhibits a symmetric nodeless pattern oriented along the $a$-axis (Figs.~\ref{Fig3}d,h). Both of M and P preserve the $C_2$ symmetry, in contrast with the observed asymmetry in MP1 and MP2. We then repeat the polarimetry measurements at a high $E_\mathrm{THz}$ ($\sim$150 kV/cm peak electric field), where M and P retain their equivalent polarimetry compared to their low $E_\mathrm{THz}$ cases (Figs.~\ref{Fig3}k,l,o,p). On the other hand, remarkable change in the polar pattern of MPs can be observed: MP1 develops two additional lobes along $\phi=\pm 135^\circ$ (Figs.~\ref{Fig3}i,m), while MP2 maintains two lobes but with increased asymmetry (Figs.~\ref{Fig3}j,n).

\paragraph{}
To quantitatively understand the distinct symmetries of the polar patterns, we fit the nonlinear and linear excitation channels based on the symmetry of the the Raman phonons and the magnon, respectively. The point group $2/m$ allows Raman-active phonons with $A_g$ and $B_g$ irreducible representations (irreps). Based on the symmetry constraints, the nonlinear and linear driving forces must share the same irrep as the corresponding phonon mode (see Supplementary Note 7). To this end, we expect the following $\phi$-dependence of the driven amplitudes of $A_g$ and $B_g$ phonons from nonlinear channels:
\begin{equation}
\label{nonlinear}
\begin{split}
    A_{\text{NL},A_g} &= a_1 E_a^2 + a_2 E_b^2  = (a_1 \cos^2\phi + a_2 \sin^2\phi)E_\mathrm{THz}^2,\\
    A_{\text{NL},B_g} &= 2a_3 E_aE_b = 2a_3 \cos\phi\sin\phi E_\mathrm{THz}^2,
\end{split}
\end{equation}
where $E_a$ and $E_b$ are the THz electric fields along the crystallographic $a$- and $b$-axes, respectively, and $a_1$, $a_2$, $a_3$ are fitting coefficients (Supplementary Note 7). For the linear channels, since the magnetization along $a$- and $b$-axes ($M_a$ and $M_b$) respects $B_g$ and $A_g$ symmetry, respectively, the $\phi$-dependence of the linear excitation channel of MP can be described as: 
\begin{equation}
\label{linear}
\begin{split}
    A_{\text{L},A_g} &\propto M_b\propto B_\mathrm{THz}\cos\phi= b_1 E_\mathrm{THz}\cos\phi,\\
    A_{\text{L},B_g} &\propto M_a\propto B_\mathrm{THz}\sin\phi= b_2 E_\mathrm{THz}\sin\phi,
\end{split}
\end{equation}
where $B_\mathrm{THz}$ is the magnetic field of the driving THz pulse, which is perpendicular to the electric field, and $b_1$, $b_2$ are fitting coefficients (Supplementary Note 7). The $C_2$ symmetry breaking behavior of MP1 and MP2 strongly suggests that they are simultaneously driven by both channels and should therefore be assigned as MPs. We can quantitatively fit the polar patterns of MP1 and MP2 by the sum of Eq.~\ref{nonlinear} and Eq.~\ref{linear} (violet lines in Figs.~\ref{Fig3}e, f, m, and n) of $B_g$ and $A_g$ symmetries, respectively. Moreover, the pattern change of MPs at different $E_\mathrm{THz}$ can be explained by the alteration in the relative strengths of the nonlinear and linear pathways, which can be extracted by the ratio between the maximum fit values of the nonlinear and linear polar patterns (color bars next to each polar pattern in Fig.~\ref{Fig3}). We find that the relative strengths not only depend on $E_\mathrm{THz}$ but also differ between the two MPs. For MP1, the dominant excitation pathway switches from linear at low $E_\mathrm{THz}$ to quadratic at high $E_\mathrm{THz}$. On the other hand, for MP2, the linear pathway dominates within our attainable $E_\mathrm{THz}$ range. In comparison, M and P can be fit by considering only the linear $A_g$ (Eq.~\ref{linear}) and the quadratic $A_g$ pathways (Eq.~\ref{nonlinear}) at all $E_\mathrm{THz}$ values, respectively (blue and red lines in Figs.~\ref{Fig3}g, h, o, and p), indicating that they are an $A_g$ magnon and an $A_g$ phonon that do not significantly hybridize with the 3.7 THz magnons, respectively.

\begin{center}
\begin{figure}[H]
   \sbox0{\includegraphics{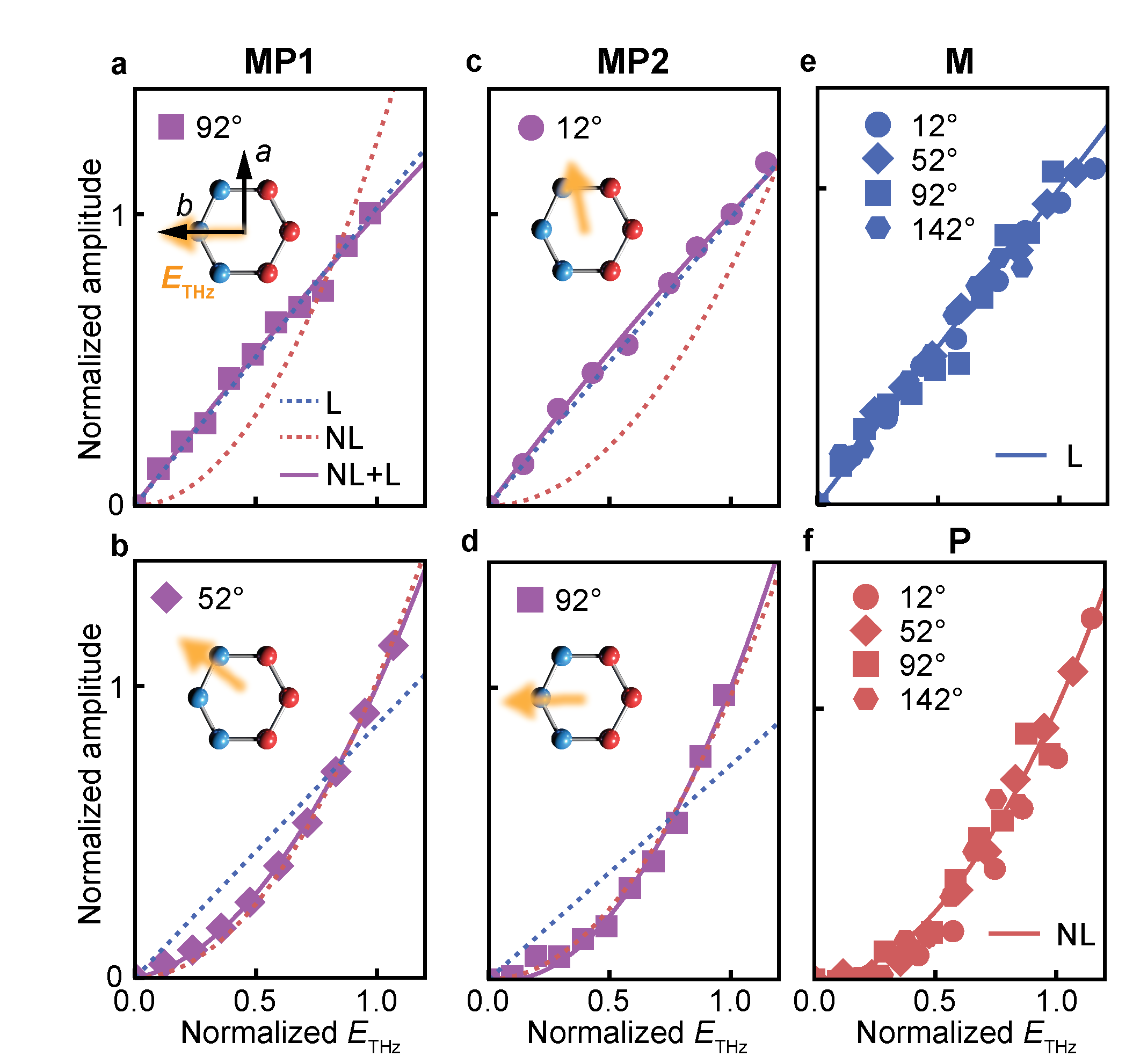}}
    \begin{minipage}{\wd0}
  \usebox0
  \captionsetup{labelfont={bf},labelformat={default},labelsep=period,name={Fig.}}
  \caption{\textbf{$E_\mathrm{THz}$ dependence of the mode amplitudes.} \textbf{a-b}, MP1 amplitude as a function of $E_\mathrm{THz}$ obtained at $\phi=92^\circ$ and $\phi=52^\circ$, respectively. \textbf{c-d}, MP2 amplitude as a function of $E_\mathrm{THz}$ obtained at $\phi=12^\circ$ and $\phi=92^\circ$, respectively. \textbf{e}, M amplitude as a function of $E_\mathrm{THz}$ at different $\phi$ values. Data are normalized for a better comparison. \textbf{f}, P amplitude as a function of $E_\mathrm{THz}$ at different $\phi$ values. Data are vertically scaled for better comparison. In all the panels, red and blue lines are quadratic and linear fits, while violet lines are the fits to Eq.~\ref{fluence}. The insets show the THz polarization with respect to the crystallographic axes. $E_\text{THz}$ is normalized by 180 kV/cm.} 
  \label{Fig2}
\end{minipage}
\end{figure}  
\end{center}

\paragraph{}
To corroborate our interpretations and directly demonstrate the existence of both linear and nonlinear pathways, we measure the $E_\mathrm{THz}$ dependencies of mode amplitudes at characteristic $\phi$ values where the linear and nonlinear responses should be nearly maximal based on the polarimetry analysis. The amplitudes of MP1 and MP2 exhibit both linear and nonlinear behaviors depending on $\phi$.  Specifically, at $\phi=92^\circ$, which is around the maximum of the linear channel (Eq.~\ref{linear}) and the node of the nonlinear channel for $B_g$ (Eq.~\ref{nonlinear}), the amplitude of MP1 shows nearly linear dependence on $E_\mathrm{THz}$ (Fig.~\ref{Fig2}a dotted blue line). The $E_\mathrm{THz}$ dependence becomes almost quadratic when we rotate the THz polarization to $\phi=52^\circ$, close to the maxima of the nonlinear channel (Fig.~\ref{Fig2}b dotted red line). For MP2, the linear and quadratic dependence on $E_\mathrm{THz}$ appear at $\phi=12^\circ$ (Fig.~\ref{Fig2}c) and $\phi=92^\circ$ (Fig.~\ref{Fig2}d), respectively, distinct from MP1. The $E_\mathrm{THz}$-dependencies of the MP amplitudes can all be quantitatively captured by Eq.~\ref{fluence}. This combined linear and quadratic  $E_\mathrm{THz}$-dependencies at distinct $\phi$ values confirm the coexistence of the two pathways. In contrast, such strong $\phi$ dependence is not observed for M and P. At all selected $\phi$ values, the amplitude of M is linear in $E_\mathrm{THz}$ (Fig.~\ref{Fig2}e), validating the magnetic-dipole excitation pathway. Similarly, the amplitude of P consistently shows a quadratic $E_\mathrm{THz}$-dependence (Fig.~\ref{Fig2}f), as expected for the sun-frequency excitation of a Raman-active phonon mode.

\begin{center}
\begin{figure}[H]
   \sbox0{\includegraphics[width=\textwidth]{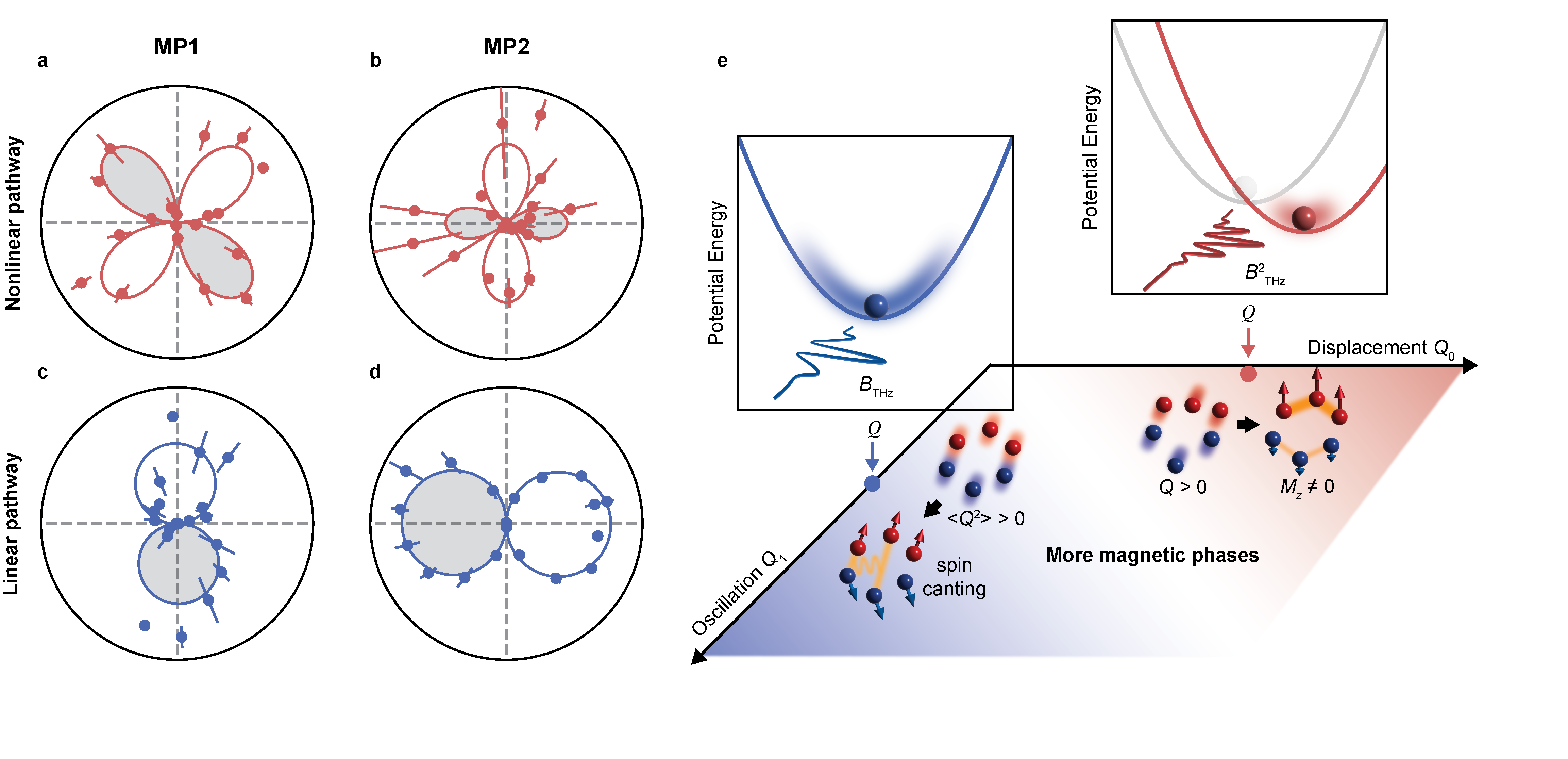}}
    \begin{minipage}{\wd0}
  \usebox0
  \captionsetup{labelfont={bf},labelformat={default},labelsep=period,name={Fig.}}
  \caption{\textbf{Nonlinear and linear pathways for MP excitation and their applications.} \textbf{a-b}, Nonlinear pathway efficiency $a$ for MP1 (\textbf{a}) and MP2 (\textbf{b}) as a function of $\phi$. Red lines are fits for the $B_g$ and $A_g$ modes to Eq.~\ref{nonlinear}. \textbf{c-d}, Linear pathway efficiency $b$ for MP1 (\textbf{c}) and MP2 (\textbf{d}) as a function of $\phi$. Blue lines are fits for the $B_g$ and $A_g$ modes to Eq.~\ref{linear}. \textbf{e}, Schematics of the dynamical phase diagram induced by coherent MPs. Red region: $M_z\neq 0$ state induced by MP2 net displacement $Q_0$. The red and blue spheres are Fe atoms with equilibrium spin up and spin down. The bonds with exchange interaction enhanced (thicker bonds) and weakened (thinner bonds) by the phonon displacement are highlighted in yellow. The red and blue arrows indicate the average spin of different Fe atoms upon THz excitation. Top inset shows the schematic of the potential energy landscape change upon driving the MP through the nonlinear displacive pathway. Blue region: spin-canting state induced by MP2 oscillation $Q_1$. The solid yellow lines depict the new spin interaction terms induced by hopping between Floquet bands  $(\mathbf{S}_i\times \mathbf{S}_j)\cdot(\mathbf{S}_k\times \mathbf{S}_l)$. Left inset shows schematic of the potential energy landscape upon driving the MP through the linear pathway, where the phonon oscillates symmetrically around the equilibrium position. White region: more magnetic phases induced by different relative strength between $Q_0$ and $Q_1$.} 
  \label{Fig4}
\end{minipage}
\end{figure}  
\end{center}

\paragraph{}
The combined measurements of $\phi$ and $E_\mathrm{THz}$ dependencies of the MP1 and MP2 amplitudes substantiate the coexistence of linear and nonlinear pathways and their interference. To further corroborate the symmetry of different MPs, we fit the $E_\mathrm{THz}$-dependence of MP amplitudes using Eq.~\ref{fluence} systematically at various $\phi$ values. We thereby obtain the $\phi$-dependence of parameters $a$ and $b$ of Eq.~\ref{fluence}, which directly reflect the symmetry of the nonlinear and linear excitation channels. Notably, $a(\phi)$ of MP1 exhibits four equal petals, each offset by 45$^\circ$ from the crystallographic axes (Fig.~\ref{Fig4}a), suggesting the $B_g$-symmetry as we previously assigned. Conversely, $a(\phi)$ of MP2 shows four petals aligned with the crystallographic axes (Fig.~\ref{Fig4}b), in agreement with the expected $A_g$-symmetry. Similarly, $b(\phi)$ of MP1 and MP2 exhibit two equal lobes aligned with the $b$- and $a$-axes (Figs.~\ref{Fig4}c and \ref{Fig4}d), respectively, consistent with the assignment of $B_g$- and $A_g$-symmetry as predicted by Eq.~\ref{linear}. Hence, the coexistence of both nonlinear and linear channels yields either constructive or destructive interference depending on the mode symmetry and $\phi$, giving rise to the ostensible $C_2$-symmetry breaking. 

\paragraph{}
To substantiate our phenomenological descriptions and quantitatively understand the experimental observations, we perform microscopic dynamical simulations (Methods). We start by constructing an equilibrium Hamiltonian describing the spin degree of freedom which includes magnetic exchange interactions and anisotropy with all the microscopic parameters determined by first-principles calculations. We then derive the spin dynamics by solving the Landau-Lifshitz-Gilbert (LLG) equation based on the Hamiltonian with a time-dependent magnetic driving field. This allows us to obtain the dynamics of the magnon-induced magnetization $M(t)$, which determines the nonlinear effective force $T_Q(t)$ (see Supplementary Note 6 and 8 for more discussions) and the linear effective force from the magnetic-dipole process $gM(t)$, where $g$ is the magnon-phonon coupling strength. The dynamics of the MP can then be simulated by its equations of motion:
\begin{equation}
\label{phonon}
\ddot{Q}+2\gamma_Q\dot{Q}+\omega_Q^2Q=T_Q(t)+gM(t),
\end{equation}
where $Q$, $\gamma_Q$, and $\omega_Q$ are the displacement amplitude, damping rate, and frequency of the MP. By comparing the theoretical simulations with the experimental observations, we achieve a remarkable agreement in terms of the dependence on $E_\mathrm{THz}$ and $\phi$, confirming the validity of our proposed excitation mechanisms (Supplementary Note 8). 

\paragraph{}
Therefore, we have comprehensively demonstrated that by selecting the appropriate $\phi$ and $E_\mathrm{THz}$ values, we can achieve a nonlinear-to-linear excitation crossover. Interestingly, for a specific phonon mode that exhibits strong coupling to the spin, such tunability allows different mechanisms for phonon-mediated control and potentially a range of different magnetic phases inaccessible in equilibrium. As an outlook, we theoretically explore different scenarios by coherently exciting MP2. In the nonlinear pathway, the quadratic driving generates a unidirectional force that results in a net displacement along the phonon coordinate\autocite{Disa2021EngineeringLight}(Fig.~\ref{Fig4}e top inset), which lasts at least within the laser pulse duration. Such displacement can effectively and selectively modify exchange interactions, giving rise to a net magnetization along the spin direction (Fig.~\ref{Fig4} red region). Our frozen phonon calculations show that the phonon displacement $Q_0$ of MP2 can bilinearly couple to the magnetization along $z$ ($M_z$) and AFM order parameter ($L$): $\mathcal{H}\propto Q_0LM_z$\autocite{Ilyas2023TerahertzFePS3}. The presence of displacement will thus generate a finite magnetization in the AFM motif\autocite{Ilyas2023TerahertzFePS3,Disa2020PolarizingField}. On the other hand, the linear excitation induces a symmetric oscillation $Q_1(t)$ of the Raman-active phonon mode without net displacement (Fig.~\ref{Fig4}e left inset). This coherent periodic modulation can produce phonon-dressed side bands\autocite{Ergecen2022MagneticallyAntiferromagnet}{}, establishing a fertile playground for manipulating electronic and magnetic properties via the phonon Floquet mechanism\autocite{Hubener2018PhononMatter,Chaudhary2020Phonon-inducedSymmetries}{} (Fig.~\ref{Fig4}e blue region). Our microscopic calculations demonstrate that to the lowest order, the phonon Floquet Hamiltonian produce interactions between four nearest neighbour spins: $\mathcal{H}\propto Q_1(t)^2(\mathbf{S}_i\times \mathbf{S}_j)\cdot(\mathbf{S}_k\times \mathbf{S}_l)$, where $\mathbf{S}$ are spins at nearest neighbour sites $i,j,k,l$. These terms are only nonzero when the spins are neither parallel nor anti-parallel to each other, thereby favoring a state with canted magnetic order (Fig.~\ref{Fig4}e blue region and Supplementary Note 9). Compared to the conventional photon-based Floquet engineering of magnetism\autocite{Mentink2015UltrafastInsulators,Claassen2017DynamicalInsulators,Kitamura2017ProbingLaser,Liu2018FloquetTitanates}{}, the extremely long lifetime (Supplementary Note 6) of the MPs in our sample allows for a sustained Floquet engineering effect that persists for over 100 ps. Finally, the crossover between nonlinear and linear excitation controlled by THz polarization $\phi$ and field strength $E_\mathrm{THz}$ tips the intricate balance between different ground states with distinguished magnetic structures, allowing for the access to a rich landscape of magnetic phases without equilibrium analogues.

\paragraph{}
Altogether, we have not only uncovered a linear pathway for resonantly exciting a Raman-active phonon through its coupling to a magnon, but also demonstrated the ability to manipulate the relative strength of the linear and nonlinear excitation routes by adjusting the strength and polarization of the driving field. Such capability enables the control over the dynamical symmetry of MPs and significantly expands the horizon of magnon- and phonon-mediated control over material properties on demand, a burgeoning field known as magno-phononics. Furthermore, the application of the methodology developed in this work to the recently discovered chiral MPs in a similar compound $\mathrm{FePSe_3}$ and other systems with strong magnon-optical phonon interactions\autocite{Torrance1972MicroscopicFeCl22H2O,Allen1971MagneticInteraction}{} affords an exceptional avenue for unprecedented methods to manipulate phonons and spins\autocite{Cui2023ChiralityAntiferromagnet,Luo2023EvidenceLimit}{}. We also envision that further nonlinear THz spectroscopy measurements on $\mathrm{FePS_3}$ will provide conclusive insights into the microscopic nature of the two-particle excitation pathways for different modes\autocite{Lopez2024Electro-phononicConversion} (Supplementary Note 6).

\section*{Methods}

\paragraph{Sample preparation\\} 
$\mathrm{FePS_3}$ single crystals were synthesized from iron (Sigma-Aldrich, 99.99\% purity), phosphorus (Sigma-Aldrich, 99.99\%), and sulfur (Sigma-Aldrich, 99.998\%) using the chemical vapor transport method. The powdered elements were prepared inside an argon-filled glove box. We weigh the starting materials in the correct stoichiometric ratio and added an additional 5 wt of sulfur to compensate for its high vapor pressure. We carried out the chemical analysis of the synthesized single-crystal samples using a COXEM-EM30 scanning electron microscope equipped with a Bruker QUANTAX 70 energy-dispersive X-ray system to confirm the correct stoichiometry. The crystal  structure of the sample is checked with a commercial X-ray diffractometer (Rigaku Miniflex II). Single crystals were cleaved along the [001] direction immediately before the experiment. The sample thickness after exfoliation is 20~$\mathrm{\mu m}$ as measured by a Bruker Dektak DXT-A stylus profilometer.

\paragraph{THz experiment\\}
A detailed description can be found in Supplementary Note 2. The broadband THz pump is generated by pumping N-benzyl-2-methyl-4-nitroaniline (BNA) crystal with the 1300 nm output from an optical parametric amplifier (OPA) seeded by a Ti:Sapphire amplifier at a repetition rate of 1 kHz. A weak 800 nm pulse from the Ti:Sapphire amplifier is focused on the sample to probe the phonon and magnon dynamics through the light-induced change in linear birefringence. The THz and 800 nm pulses are incident normally on the sample. The ellipticity of the 800 nm pulse after transmitting through the sample is measured by the balanced detection scheme with a pair of photodiodes. The output of the photodiodes is collected by a data acquisition (DAQ) card.

\paragraph{Theoretical simulations} 
The microscopic Hamiltonian describing the spin degree of freedom of the system can be expressed as:
\begin{equation}
    \mathcal{H}=J_1\sum_{{\langle}ij{\rangle}} \mathbf{S}_i\cdot\mathbf{S}_j+J_2\sum_{{\langle\langle}ij{\rangle\rangle}} \mathbf{S}_i\cdot\mathbf{S}_j+J_3\sum_{{\langle\langle\langle}ij{\rangle\rangle\rangle}}\mathbf{S}_i\cdot\mathbf{S}_j-\Delta\sum_{i} S_{iz}^2-\gamma \mathbf{B}\cdot\sum_{i}\mathbf{S}_i,
    \label{Hamiltonian}
\end{equation}
where $\mathbf{S}_i$ is the spin of Fe at site $i$. The first three terms represent Heisenberg interactions between nearest, next-nearest, and next-next-nearest in-plane neighbors, the fourth term corresponds to an Ising type easy-axis anisotropy, and the last term is the Zeeman energy due to an in-plane magnetic field which represents the THz driving field. Interactions between layers can be neglected. $\gamma=g\mu_B/\hbar$ is the gyromagnetic ratio. The values of all exchange interactions and anisotropies have been obtained through first-principles calculations\autocite{Zhang2021CoherentInsulator,Ilyas2023TerahertzFePS3}{}. Then we solve the Landau–Lifshitz–Gilbert (LLG) equation, which describes the dynamics of the motion of spins under an effective magnetic field $\mathbf{H}_\text{eff}=-\frac{\partial\mathcal{H}}{\gamma \partial\mathbf{S}_i}$. It can be expressed as:
\begin{equation}
    \frac{d\mathbf{S}_i}{dt}=\frac{\gamma}{1+\alpha^2}[\mathbf{S}_i\times\mathbf{H_{eff}}+\frac{\alpha}{|\mathbf{S}_i|}\mathbf{S}_i\times(\mathbf{S}_i\times\mathbf{H}_\text{eff})],
    \label{LLG}
\end{equation}
where $\gamma=g\mu_B/\hbar$ is the gyromagnetic ratio and $\alpha$ is a phenomenological Gilbert damping constant that accounts for energy dissipation. We can then solve the dynamical equation for the magnetization $\mathbf{M}$, summing over the four spins within one unit cell: $\mathbf{M}=\mathbf{S}_1+\mathbf{S}_2+\mathbf{S}_3+\mathbf{S}_4$. As we solve the dynamics of $\mathbf{M}$, we can study their impact on phonons through magnon-phonon coupling based on the symmetry constraints. In the linear coupling channel, $A_g$ phonon exclusively couples to $A_g$ magnon while $B_g$ phonon exclusively couples to $B_g$ magnon, which induce magnetization along $y$ ($M_y$) and $x$ ($M_x$), respectively. Thus, we have:
\begin{equation}
\begin{split}
     &\ddot{Q}_{Ag}+2\dot{Q}_{Ag}/\tau_{Ag}+\omega_{Ag}^2Q_{Ag}=g_{Ag}M_y,  \\
     &\ddot{Q}_{Bg}+2\dot{Q}_{Bg}/\tau_{Bg}+\omega_{Bg}^2Q_{Bg}=g_{Bg}M_x,
\end{split}
\label{PhononLinearDynamics}
\end{equation}
where $Q_{Ag/Bg}$, $\tau_{Ag/Bg}$, $\omega_{Ag/Bg}$, and $g_{Ag/Bg}$ are the phonon coordinate, phonon lifetime, phonon frequency, and phonon-magnon coupling constant of $A_g$ and $B_g$ symmetry phonons, respectively. For the nonlinear driving, we consider the one-magnon-one-photon excitation pathway. As discussed in Supplementary Note 7, other quadratic driving pathways should produce qualitatively similar $E_\mathrm{THz}$ and $\phi$ dependence. Based on the symmetry, we have:
\begin{equation}
\begin{split}
     &\ddot{Q}_{Ag}+2\dot{Q}_{Ag}/\tau_{Ag}+\omega_{Ag}^2Q_{Ag}=g_{Ag1}M_yB_y+g_{Ag2}M_xB_x,  \\
     &\ddot{Q}_{Bg}+2\dot{Q}_{Bg}/\tau_{Bg}+\omega_{Bg}^2Q_{Bg}=g_{Bg1}M_yB_x+g_{Bg2}M_xB_y,
\end{split}
\label{PhononNonLinearDynamics}
\end{equation}
where $g_{Ag/Bg}$ are the nonlinear phonon-magnon coupling constants of $A_g$ and $B_g$-symmetry phonons, respectively.
The simulation results of LLG simulations are elaborated in Supplementary Note 8.

\section*{Data Availability}
\noindent The datasets generated and/or analysed during the current study are available from the corresponding author on request.

\section*{Code Availability}
\noindent The code used for the current study are available from the corresponding author on request.

\printbibliography

\section*{Acknowledgments}
\noindent We thank Zhuquan Zhang and Keith Nelson for help with the experiments. We acknowledge the support from the US Department of Energy, Materials Science and  Engineering  Division,  Office  of  Basic  Energy  Sciences  (BES  DMSE)  (data taking and analysis), Gordon and Betty Moore Foundation’s EPiQS Initiative grant GBMF9459 (instrumentation and manuscript writing), and the MIT-Israel Zuckerman STEM Fund. E.V.B. acknowledges funding from the European Union's Horizon Europe research and innovation programme under the Marie Skłodowska-Curie grant agreement No 101106809. A.R. was supported by the Cluster of Excellence Advanced Imaging of Matter (AIM), Grupos Consolidados (IT1249-19), SFB925, “Light Induced Dynamics and Control of Correlated Quantum Systems,” and the Max Planck Institute New York City Center for Non-Equilibrium Quantum Phenomena. A.v.H. gratefully acknowledges funding by the Humboldt Foundation. D.M.J. is supported by Tel Aviv University. The work at SNU was supported by the Leading Researcher Program of Korea’s National Research Foundation (Grant No. 2020R1A3B2079375).

\section*{Author Contributions}
\noindent T.L., H.N. and B.I. performed the measurements. H.N. performed dynamical simulations with help from E.V.B. E.V.B. and A.R. developed the theoretical model of phonon Floquet effect. J.P. and J.K. synthesized and characterized \ce{FePS3} single crystals under the supervision of J.-G.P. T.L., H.N., A.v.H. and B.I. performed the data analysis. T.L., H.N., B.I., A.v.H., D.M.J., E.V.B. and N.G. interpret the data and wrote the manuscript with critical inputs from all other authors. The project was supervised by N.G.

\section*{Competing Interests}
\noindent The authors declare no competing interests.

\newpage

\newpage

\newpage

\newpage

\mbox{~}
\clearpage
\end{document}